\documentclass{ws-ijmpa}

\begin{document}
\markboth{V. Yu. Lazur, O. K. Reity \& V. V. Rubish}
{Spherical model of the Stark effect in external scalar and vector fields}

\catchline{}{}{}{}{}
%
%%%%%%%%%%%%%%%%%%%%%%%%%%%%%%%%%%%%%%%%%%%%%%%%%%%%%%%%%%%%%%%%%%%%

\title{SPHERICAL MODEL OF THE STARK EFFECT \\IN EXTERNAL SCALAR AND VECTOR FIELDS}

\author{V.~Yu.~LAZUR$^{*}$, O.~K.~REITY$^{\dag}$ and V.~V.~RUBISH$^{\ddag}$}

\address{Department of Theoretical Physics, Uzhgorod National University, \\Voloshyna Street 54, Uzhgorod 88000, Ukraine\\
$^{*}$lazur@univ.uzhgorod.ua
\\$^{\dag}$reiti@univ.uzhgorod.ua
\\$^{\ddag}$vrubish@univ.uzhgorod.ua}

\maketitle

\begin{history}
\received{Day Month Year}
\revised{Day Month Year}
\end{history}

\begin{abstract}
The Bohr-Sommerfeld quantization rule and the Gamow formula for
the width of quasistationary level are generalized by taking
into account the relativistic effects, spin and Lorentz structure
of interaction potentials. The relativistic quasi-classical
theory of ionization of the Coulomb system ($V_C=-\xi/r$) by
radial-constant long-range scalar ($S_{l.r.}=(1-\lambda)(\sigma
r+V_0)$) and vector ($V_{l.r.}=\lambda(\sigma r+V_0)$) fields is
constructed. In the limiting cases the approximated analytical
expressions for the position $E_r$ and width $\Gamma$ of below-barrier
resonances are obtained. The strong dependence of the width
$\Gamma$  of below-barrier resonances on both the bound level energy
and the mixing constant $\lambda$ is detected. The simple
analytical formulae for asymptotic coefficients of the Dirac
radial wave functions at zero and infinity are also obtained.

\end{abstract}

\ccode{PACS numbers: 03.65.Pm, 03.65.Sq, 03.65.Xp, 31.15.Gy,
32.60.+i, 32.70.Jz}

\section{Introduction}\label{s1}

Wide range of problems from various fields of physics (elementary particle physics, nuclear physics, physics of atomic collisions, etc.) is related to representations of formation and decay of non-stable (quasistationary) states of quantum systems.\cite{Baz} Properties of such states are of interest for investigation of ionization of atoms, ions and semiconductors under the influence of constant and homogeneous electric and magnetic fields,\cite{Popov_Tunel} for description of cluster decays of atomic nuclei\cite{Goryachev} and effects of the spontaneous creation of positrons\cite{Popov_Rev,Zeldovich}, in consideration of a vacuum shell of supercritical atom,\cite{Popov_Rev}\cdash\cite{Grib} and from the point of view of studying the Dirac equation in the strong external fields as well.

The relativistic theory of decaying (quasistationary) states is elaborated quite well for the cases when components of an interaction potential of a fermion with external fields belong to the vector type, i.e. are Lorentz-vector $A_{\mu}$ components.
\cite{Popov_Tunel,Zeldovich}\cdash\cite{Greiner} At the same time the interior logic of development of the decaying states theory obviously dictates the statement of qualitatively new problems having an origin in nuclear physics and elementary particle physics. So, for example, quark structure of nucleons and multi-quark states in a nucleus compel to look at the nature of the inter-nuclear forces in a new way as well. Under the stimulating influence of QCD the problem of manifestation of quark-gluon degrees of freedom in atomic nuclei and nuclear processes has gained new development in the last decade and now, undoubtedly, constitutes the main prospect of fundamental researches in this field of nuclear physics. Without going into the further discussion of these important problems, note only that there are already enough reviews, books and papers where the specified tendency is sufficiently reflected (see, for example Ref.~\refcite{Sitenko}, and references therein).

As is known, the problem of penetration through potential barriers underlies a theoretical view of the phenomena of cluster decays of atomic nuclei and non-stable resonance states of strongly interacting elementary particles. The subsequent theoretical description of such phenomena should be based on the relativistic wave equations by taking into account that besides electromagnetic forces the interactions between elementary particles can be also realized by the forces which are not dependent on the electrical charge. Take for example nuclear interactions between the nucleons, caused by interaction of nucleons with the meson field. One more class of (this time long-range) forces related to the considered problem are the forces arising between nucleons during exchange of electron-neutrino pairs (so-called $\beta$-forces). For the first time such forces have been introduced by I. Tamm\cite{Tamm} in 1934, and more recently the vector and pseudovector variants of these interactions were considered in detail by authors of the well-known monography.\cite{Greiner} From the point of view of new problems of the strong interaction theory it is interesting to explore the more general case when a spin-1/2 particle interacts with the scalar and vector fields simultaneously. As is known now there is a reason to think that such interactions exist between quarks in hadrons.

The main difficulties of the theory of quasistationary states (for the applications specified above) are caused by the fact that in many cases the interactions cannot be described within the standard methods that use expansions in small energy parameter. In the problems related to the description of quasistationary states of the relativistic composite systems, additional difficulties arise when solving the Dirac equation with unseparable variables. In the modern theory of decaying states these difficulties are overcome by means of the relativistic version of the imaginary time method developed in papers.\cite{Popov_Tunel,Mur_MMB} This allows to calculate the tunneling probability of relativistic particles through potential barriers including those that do not possess spherical symmetry.

Though this method has heuristic force and physical clarity, nevertheless it cannot be considered to be strictly mathematically justified, despite some attempts that were made in this direction.\cite{Popov_Tunel,Marinov} As is known, the accounting of the Coulomb interaction between the outgoing electron and atomic core within the imaginary time method encounters considerable difficulties and, for example, in the theory of multiphoton ionization of atoms\cite{Popov_Tunel} is not fully carried out up to present time.

Fortunately, many interesting questions of the relativistic theory of quasistationary states can be elucidated on an example of the simple models permitting exact or asymptotically exact solution of the Dirac equation. From the variety of such problems here we consider the hybrid version (\ref{2}) of \emph{spherical model of the Stark effect} (SMSE). Inclusion in the standard SMSE of ``new'' interactions, related to the scalar field, opens new possibilities for its applications in the relativistic nuclear physics and QCD. In a more general context we consider the nonstandard modeling problem of study of simultaneous influence of the radial-constant scalar and vector fields on a system of Coulomb levels.

The quasistationary solutions of the Dirac equation in the composite field (\ref{2}) at $1/2<\lambda\leqslant 1$ and the corresponding complex spectrum of energy are generated by the radiation requirement meaning that at infinity the solutions $F(r)$, $G(r)$ are the divergent waves (see Ref.~\refcite{part1}). If $E$ is real, such solution does not exist, however, there are infinitely many complex quasistationary levels $E=E_r-i\Gamma/2$ with exponentially small imaginary part $\Gamma$.

For the field (\ref{2}) consisting from the mixture of Coulomb and radial-constant vector and scalar fields, the Dirac equation is separable in spherical coordinates. Apparently, this circumstance should essentially facilitate calculation of the position $E_r$ and width $\Gamma$ of resonance. Let us note, however, that the usual approach (the numerical solution of the Dirac system) encounters the known difficulties related to exponential increasing of the Gamow wave function (at $r\rightarrow\infty$) of a quasistationary state. In view of complexity of this problem we shall solve it in the quasi-classical approximation that gives the useful analytical expressions for the position $E_r$ of resonance and its width $\Gamma$. Except the reasons of convenience the WKB method, or quasi-classical approximation, possesses a number of the principle advantages as compared to other methods.
As is known, unlike the perturbation theory this approach is not related to the smallness of interaction and, consequently, has a wider range of applicability, allowing to investigate qualitative regularities in behavior and properties of quantum mechanical systems. Other important advantage of this method is in its applicability to the cases of both electromagnetic and scalar external fields. Further we shall consider the version of WKB method offered in Refs.~\refcite{part1,Lazur} that can be used in the case of both the discrete spectrum and quasistationary states (resonances).

This paper is organized as follows. In Sec.~2 we
generalize the Bohr–Sommerfeld quantization rule to the relativistic case where a spin-1/2 particle interacts
with scalar and electrostatic external fields simultaneously. In the following section by means of WKB method we solve the problem of finding asymptotic coefficients of the wave function in zero and at infinity with appropriate accuracy. Sec.~4 is devoted to the construction of the quasi-classical theory of ionization of Coulomb system ($V_C=-\xi/r$) by radial-constant scalar ($S_{l.r.}=(1-\lambda)(\sigma
r+V_0)$) and vector ($V_{l.r.}=\lambda(\sigma r+V_0)$) fields by taking into account the relativistic effects and fermion spin. In the limiting cases $\sigma/\xi\tilde {m}^2\ll 1$ and $\sigma\gamma/\widetilde{E}_r^2\ll 1$ the approximate analytical expressions for width of below-barrier resonances $\Gamma$ which show the strong dependence of $\Gamma$ on both the energy of bound level $E_r$ and mixing parameter $\lambda$ are obtained.

\section{Position of quasistationary states}

Separating the angular variables in the Dirac equation with the spherically symmetric vector $V(r)$ and scalar $S(r)$ interaction potentials, we obtain the system of first-order ordinary differential equations for the radial wave functions $F$ and $G$ ($c=1$):
\begin{equation}\label{1}
\left.\begin{array}{c}
\displaystyle\hbar\frac{dF}{dr}+\frac{\tilde{k}}{r}F-
\left[\left(E-V(r)\right)+\left(m+S(r)\right)\right]G=0,\\
\displaystyle\hbar\frac{dG}{dr}-\frac{\tilde{k}}{r}G+\left[\left(E-V(r)\right)-
\left(m+S(r)\right)\right]F=0.\end{array}\right\}
\end{equation}
Here $\hbar$ is the Planck constant, $\tilde{k}=\hbar\,k$,
$k=\mp(j+1/2)$ is the motion integral of the Dirac particle in a central field, $j\hbar$ is the total angular moment; the definition and normalization of the function $F$ and $G$ are the same as in the papers.\cite{part1,Lazur}

\begin{figure}[pb]
\centerline{\psfig{file=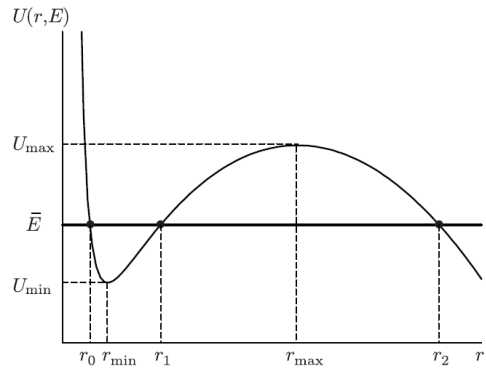,width=75mm}}
\vspace*{8pt}
\caption{Barrier-type effective potential $U(r,E)$; $r_0$, $r_1$ and $r_2$ are roots of the equation $p^2(r)=0$.}\label{f1}
\end{figure}

For the description of the phenomena related to the formation and decay of quasistationary states, we consider the class of potentials $V(r)$ and $S(r)$ for which the effective potential (see (\ref{3})) of the squared Dirac equation possesses a barrier (of the type shown in Fig.~\ref{f1}).

Let us give the algorithm of construction of quasi-classical solutions of the Dirac system (\ref{1}). As usually, we seek a solution of this system in the form of product of rapidly oscillating phase factor and slowly oscillating amplitude:
\begin{eqnarray*}
  \chi(r)&=&\left(\begin{array}{c}F(r)\\
G(r)\end{array}\right)=\exp
\left\{\int\limits^{r}y(r')\, dr' \right\}\varphi(r), \\
 y(r)&=&\hbar^{-1}y_{-1}(r)+y_{0}(r)+\hbar\,y_{1}(r)+\ldots,\quad
\varphi(r)= \sum^{\infty}_{n=0}\hbar^{n}\varphi^{(n)}(r),
\end{eqnarray*}
where $\hbar\rightarrow 0$ is a small parameter, $\varphi(r)$ and
$\varphi^{(n)}(r)$ are the two-component quantities (the upper and lower components of $\varphi^{(n)}(r)$ correspond to the radial functions $F$ and $G$ respectively).
Expansion of $y(r)$ and $\varphi(r)$ in $\hbar$ leads to a chain of the matrix differential equations for $y_{(n)}(r)$ and $\varphi^{(n)}(r)$ which are solved consequentially by means of known technic of the left and right eigenvectors of the homogeneous system. Let us give the final formulae for the wave function of a quasistationary state. These formulae have a different view in three regions: i) a potential well $r_{0}<r<r_1$; ii) below-barrier region $r_{1}<r<r_2$; iii) classically allowed region $r>r_{2}$ with quasi-discrete energy spectrum. Here $r_0$, $r_1$ and $r_2$ are the turning points, in which the radial quasi-classical momentum
\[
p(r)=\sqrt{(E-V(r))^{2}-(m+S(r))^{2}-(k/r)^2}
\]
becomes zero (see Fig.~\ref{f1}). In the classically  allowed region $r_{0}<r<r_1$ we have:
\begin{eqnarray}
  F(r)&=&C_{1}^{\pm}\left(\displaystyle\frac{E-V+m+S}{p(r)}\right)^{1/2}\cos
  \Theta_{1},\nonumber\\
  G(r)&=&C_{1}^{\pm}\mathrm{sgn}\,k\left(\displaystyle\frac{E-V-m-S}{p(r)}\right)^{1/2}\cos
  \Theta_{2}.\label{2a}
\end{eqnarray}
Here we use the following new notations
\begin{eqnarray}\label{2c}
&\displaystyle\Theta_{1}(r)=\int\limits^{r}_{r_{1}}\left(p+\frac{k\,w}{p\,r}\right)dr+\frac{\pi}{4},
    \quad
\Theta_{2}(r)=\int\limits^{r}_{r_{1}}\left(p+\frac{k\,\tilde{w}}{p\,r}\right)dr+
    \frac{\pi}{4},\\
&\displaystyle\label{2d}
w=\frac{1}{2}\left(\frac{V'-S'}{m+S+E-V}-\frac{1}{r}\right), \quad
\tilde{w}=\frac{1}{2}\left(\frac{V'+S'}{m+S-E+V}+\frac{1}{r}\right).
\end{eqnarray}
If a width of quasistationary level $ \Gamma\ll 1$ (which is justified by the answer) the condition of normalization by a single particle localized in the region $r_{0}<r<r_1$ defines normalization constants $C_{1}^{\pm}$:
\begin{equation}\label{t}
\int\limits^{r_{1}}_{r_{0}}\left(F^2+G^2\right)dr=1,\qquad
\left|C_{1}^{\pm}\right|=\left\{\int\limits^{r_{1}}_{r_{0}}\frac{E-V(r)}{p(r)}dr\right\}^{-1/2}=
\left(\frac{2}{T}\right)^{1/2},
\end{equation}
where $T$ is the period of radial oscillations of a classical relativistic
particle localized in the region I ($r_{0}<r<r_1$).

In the below-barrier region ($r_{1}<r<r_2$) at $k>0$ the solution, corresponding to the decreasing exponent, is of the form
\begin{eqnarray}
\chi&=&\left(
\begin{array}{c}
F\\G
\end{array} \right)\nonumber\\
&=&\displaystyle\frac{C^{+}_{2}}{\sqrt{qQ}}\exp\left\{-\int\limits^{r}_{r_{2}}
\left[q+\frac{(m+S)V'+(E-V)S'}{2\,qQ}\right]dr\right\}\left(
\begin{array}{c}
-Q \\ m+S-E+V \\ \end{array}\right),\label{2e}
\end{eqnarray}
and for states with $k<0$
\begin{equation}\label{2f}
\chi=\frac{C^{-}_{2}}{\sqrt{qQ}}\exp\left\{-\int\limits^{r}_{r_{2}}
\left[q-\frac{(m+S)V'+(E-V)S'}{2\,qQ}\right]dr\right\}\left(
\begin{array}{c}
m+S+E-V \\ -Q \\ \end{array} \right).
\end{equation}
Here $q=|p(r)|$, $Q=q+|k|r^{-1}$.

At last, at $r>r_2$ the divergent wave corresponds to the quasistationary state;  at $k> 0$ the quasi-classical formulae for $F$ and $G$ have the form
\begin{equation}\label{2k}
\chi=\frac{C^{+}_{3}}{\sqrt{pP}}\exp\left\{\int\limits^{r}_{r_{2}}
\left[ip+\frac{(m+S)V'+(E-V)S'}{2\,pP}\right]dr\right\}\left(
\begin{array}{c}
      iP \\ m+S-E+V \end{array} \right),
\end{equation}
where $P=p+i|k|r^{-1}$. For states with $k<0$ the radial wave function $\chi$ is given by
\begin{equation}\label{2l}
\chi=\frac{C^{-}_{3}}{\sqrt{pP}}\exp\left\{\int\limits^{r}_{r_{2}}
\left[ip-\frac{(m+S)V'+(E-V)S'}{2\,pP}\right]dr\right\}\left(
\begin{array}{c}
      m+S+E-V \\iP\end{array}\right).
\end{equation}

The quasi-classical representations (\ref{2a})-(\ref{2l}) constructed are invalid in small neighbourhoods of turning points $r_j$
($j=0,1,2$). To bypass these points and match the solutions one can use the Zwaan method\cite{Zwaan} that allows to establish relations between normalization constants:
\begin{eqnarray}\label{2m}
&\displaystyle C_{2}^{\pm}=-i\,C_{3}^{\pm}=\mp
\frac{C_{1}^{\pm}}{2}
\left[\frac{E-V\left(r_1\right)+m+S\left(r_1\right)}
{\left|k\right|r_1^{-1}}\right]^{\pm\frac{1}{2}} \nonumber \\
&\displaystyle\times\exp
\left\{-\int\limits_{r_{1}}^{r_{2}}\left[q \pm
\frac{\left(m+S\right)V'+\left(E-V\right)S'}{2\,qQ}\right]dr\right\}.
\end{eqnarray}

The quasi-classical formulae obtained above allow to solve a wide range of problems in the theory of quasistationary states. So, in the case of exponential smallness of the barrier penetrability in the effective potential (EP)
\begin{equation}\label{3}
U\left(r,E_r\right)=\frac{E_r}{m}V+S+\frac{S^{2}-V^{2}}{2m}+\frac{k^{2}}{2m\,r^{2}},\quad
\bar{E}_r=(E_r^2-m^2)/2m
\end{equation}
the real part of level energy $E_r=E_{n_r k}$ is defined by the quantization condition
\begin{equation} \label{3a}
\int\limits_{r_{0}}^{r_{1}}\left(p+\frac{k\,w}{p\,r}\right)dr=
\left(n_{r}+\frac{1}{2}\right)\pi, \qquad n_r=0,1,2,\ldots,
\end{equation}
where $n_r$ is the radial quantum number.

Now we explore one concrete example of the vector and scalar potentials
\begin{equation}\label{2}
V(r)=-\frac{\xi}{r}+\lambda v(r), \quad S(r)=(1-\lambda) v(r),
\quad v(r)=\sigma r+V_0,
\end{equation}
in whose mixture at $1/2<\lambda\leqslant1$ and any $\sigma\neq 0$ quasistationary states of the Fermi particles exist; here $V_0$ is a real constant, $\xi$ is the Coulomb coupling constant, and $\lambda$ is the coefficient of mixing between the vector and scalar parts of the long-range potential $v(r)$. The potential $v (r)$ includes contributions of Lorentz-vector $V_{l.r.}(r)=\lambda v(r)$ and Lorentz-scalar $S_{l.r.}(r)=(1-\lambda)v(r)$ components, of which (at $1/2<\lambda\leqslant1$) the first one dominates in all range of $r$, $0<r<\infty$.

Even in the case of the relatively simple model of an interaction (\ref{2}) the effective potential $U\left(r,E_r\right)$ (\ref{3}) is a complicated function of the initial potentials $V(r)$ and $S(r)$, level energy $E_r$, total moment $j$, and has an essentially different shape at $0\leqslant\lambda<1/2$, $1/2<\lambda\leqslant1$ and $\lambda=1/2$. The appearance of the quasistationary states in the potential model (\ref{2}) at $1/2<\lambda\leqslant1$ and $\sigma\neq 0$ is very natural. Under specified conditions at small distances ($r\lesssim 1$) EP $U\left (r, E_r\right)$ corresponds to repulsion proceeding to the quadratic attraction at $r\gg 1$. It indicates at the definitive role of the long-range vector forces (electrostatic ones for example) in a barrier occurrence in EP $U\left(r,E_r\right)$. This ensures a possibility of level decay by particle penetration through the potential barrier in EP $U\left(r,E_r\right)$. Thus, at $1/2<\lambda\leqslant 1$ the bound state is transformed into the quasistationary one. At $r\rightarrow\infty$ its wave function has an asymptotic behavior of of a divergent wave type (\ref{2k}), (\ref{2l}). At the same time, at any $\sigma\neq 0$ the energy $E(\sigma)$ becomes complex ($E=E_r-i\,\Gamma/2$), and its imaginary part is directly connected to the probability of level ionization by the external field (\ref{2}).

Interest to the potential model (\ref{1}), (\ref{2}) at 0$ \leqslant\lambda <1/2$ is basically caused by its use for the description of mass spectra of  mesons and baryons and is not exhausted even till now (see, for example Refs.~\refcite{part1,Mur4}, and references therein). It successfully combines the effective Coulomb interaction at small distances $r$ with the linear increasing at large $r$ that leads to a confinement of color quarks and corresponds to the string ``stretched'' between quarks. In our recent paper\cite{part1} the detailed calculations of spectra of heavy-light mesons were carried out by means of the potentials (\ref{2}). In this case the model parameters (\ref{2}) are determined by a standardly: $\sigma$ is the string tension, $\xi$ is the Coulomb coefficient, $V_0$ is eigenenergy of a static source, and parameter $\lambda$ ($0\leqslant\lambda<1/2$) gives the relative contribution of the vector $V_{l.r.} (r) $ and scalar $S_{l.r.}(r)$ potentials.

The term $(S^2-V^2)/2m $ in EP (\ref{3}) dominates at large distances and at $0\leqslant\lambda <1/2$ leads to the effective repulsion regardless of the sign of the parameter $\sigma$. This repulsion is purely a relativistic effect and related to the fact that the interaction of fermion with the scalar external field $S(r)$ is added to the scalar quantity $m$ (particle mass), whereas the vector potential $V(r)$ is introduced into the free Dirac equation by the minimum way as the temporal component of the 4-potential
$A_\mu$. The presence of the relativistic terms $(S^2-V^2)/2m$ in EP $U\left(r, E_r\right)$ leads to the fact that $U\left(r,E_r\right)$ ($\sim(1-2\lambda)(\sigma r)^2/2m $) increases quadratically at large distances when $0\leqslant\lambda <1/2$. This ensures a confinement of color quarks and gluons (not observed in the free state). These properties of EP $U\left(r,E_r\right)$ (\ref{2}) of the interaction model indicate that namely scalar fields, not vector ones, are essential for confinement.

Considering the inter-quark potential, it is worthwhile to investigate the qualitative regularities of the quasi-classical spectrum of the interaction model (\ref{2}) in the phase of outgoing color, $1/2 <\lambda\leqslant1$. One can even say that without such study it is impossible to completely understand the physics of confinement.

Of course, the considered physical objects (quarks) cannot be free particles. However, it is not necessary to imagine that the string between quarks can become infinitely long.
Outside of applicability of perturbation theory, at large distances the strong coupling inevitably leads to the creation of color charged particles and the problem becomes multiparticle in essence.
Therefore, the problem about scattering of quarks, forming such colorless objects as hadrons, is usually considered. The same is valid for QCD problems related to study of quasistationary states of potential type. Here the problems of influence of the most various physical factors such as external fields with the mixed Lorentz-structure of interaction potentials, adiabatically slow variation of interaction parameters, connections with other (exterior and interior) degrees of freedom, translation and spin symmetry etc. on properties of decaying states arise quite naturally.

Basically, one can not exclude that the results obtained when solving similar problems may appear useful for the description of tunnel effects in physics of condensed matter (for example, in two-zonal semiconductors). So, for example, the electron in its motion in a solid behaves as a particle with ``effective'' mass $m_{eff}$, which differs considerably from the true mass $m$. Moreover, $m_{eff}$ can depend on distance $r$ and motion features, because the difference of masses ($m_{eff}-m$) is actually caused by interaction of an electron with objects surrounding it. In the same sense all quarks masses are ``effective'' because they are determined in processes in which a quark interacts with other quarks.

In the following two sections we shall set out the applications of elaborated quasi-classical formalism to calculation of shifts and widths of quasistationary states of the potential model (\ref{2}) with outgoing color charges ($1/2<\lambda\leqslant1$).
Special case $\lambda=1/2$, separating a phase of confinement of color charges and a phase of outgoing color (i.e. deconfinement), we will consider in our next papers.

Following the described in Ref.~\refcite{part1} (see Sec.~4 therein) calculation scheme, we represent the quasi-classical momentum in a convenient form:
\[
p(r)=|\sigma|\sqrt{2\lambda-1}\,\frac{R_1(r)}{r}=|\sigma|\sqrt{2\lambda-1}\,
\frac{\left[(a-r)(b-r)(r-c)(r-d)\right]^{1/2}}{r}.\]
Here $ \sigma $ plays a role of intensity of the radial-constant long-range field, and the turning points $a$, $b$, $c$ and $d$ are solutions of the equation
$r^{4}+f\,r^{3}+g\,r^{2}+h\,r+l=0$ with coefficients
\begin{eqnarray}
 f=\displaystyle\frac{2\eta_1} {\left(1-2\lambda\right)\sigma},\quad
g=-\frac{\widetilde{E}_r^{2}-\widetilde{m}^{2}-2\xi\sigma\lambda}
{\left(1-2\lambda\right)\sigma^{2}},\quad
h=-\displaystyle\frac{2\widetilde{E}_r\xi}{\left(1-2\lambda\right)\sigma^{2}},\quad
l=\displaystyle\frac{\gamma^{2}}{\left(1-2\lambda\right)\sigma^{2}},\nonumber
\end{eqnarray}
where $\gamma=\sqrt{k^2-\xi^2}$ and
$\eta_1=\left(1-\lambda\right)\widetilde{m}+\lambda\widetilde{E}_r$.
The characteristic parameters $\widetilde{E}_r=E_r-\lambda
V_0$ and $\widetilde{m}=m+(1-\lambda)V_0$ introduced here mean the ``shifted'' energy and the ``shifted'' mass, respectively. The specified equation has four real roots ($d <c <b <a $), determined by the expressions
\begin{equation} \label{A2}
a,b=-\frac{f}{4}+\frac{1}{2}\left(\Xi \pm \Delta_{+}
\right),\quad\quad
 c,d=-\frac{f}{4}-\frac{1}{2}\left(\Xi \mp
\Delta_{-}\right),
\end{equation}
where
\begin{eqnarray}
\Xi&=&\left[\displaystyle\frac{f^2}{4}-\frac{2g}{3}+\frac{u}{3}\left(
\frac{2}{Z}\right)^{1/3}+\frac{1}{3}\left(\frac{Z}{2}\right)^{1/3}
\right]^{1/2},\quad \Delta_{\pm}=\sqrt {F\pm
\displaystyle\frac{D}{4\Xi}},\nonumber\\
F&=&\displaystyle\frac{f^2}{2}-\frac{4g}{3}-\frac{u}{3}\left(\frac{2}{Z}\right)^{1/3}
-\frac{1}{3}\left(\frac{Z}{2}\right)^{1/3},\quad
Z=v+\sqrt{-4u^3+v^2},\nonumber \\
D&=&-f^3+4fg-8h,\quad u=g^2-3fh+12l,\nonumber\\
v&=&2g^3-9fgh+27h^2+27f^2l-72gl,\nonumber
\end{eqnarray}
the upper signs in (\ref{A2}) correspond to the turning points $a$ and $c$, and the lower ones correspond to the points $b$ and $d$.

In the quantization condition (\ref{3a}) we integrate over the range of $r$ where $\bar{E}-U(r,E_r)>0$. For the potentials (\ref{2}) considered by us at $1/2<\lambda\leqslant 1$ this means that $r_{0}=c$ and $r_{1}=b $ where $c<r<b$ and $r>a$ are classically allowed regions, and $b<r<a$ is below-barrier region, in which $p^2(r)<0$; at $r>a$ the particle goes to infinity. If the maximum of the potential function $U(r,E_r)$ is at the point $r_{max}\approx\eta_1[(2\lambda-1)\sigma]^{-1}$ for $\sigma>0$ ($r_{max}\approx[\widetilde{E}_r\xi/(\eta_1|\sigma|)]^{1/2}$ for $\sigma<0$), and minimum is at $r_{min} \approx\gamma^2/\widetilde {E} _r\xi $ then the energy spectrum of below-barrier resonances is in the range $U_{min}<\bar{E_r}<U_{max}$. The considered situation is schematically shown in Fig.~\ref{f1} where black dots indicate the position of turning points.

Using the technique of evaluation of quantization integrals from Sec.~4 of Ref.~\refcite{part1}, we can represent the phase integrals of the equation (\ref{3a}) in  the form
\begin{equation}\label{20}  J_{1}=\int\limits_{c}^{b}p(r)dr=\sqrt{2\lambda-1}\,|\sigma|
\int\limits_{c}^{b}\frac{\left(r^{3}+fr^{2}+gr+h+lr^{-1}\right)}{R_1}dr,
\end{equation}
\begin{equation}\label{21}
 J_{2}=\int\limits_{c}^{b}\frac{k\,w}{p(r)r}dr=-\frac{k}{2\sqrt{2\lambda-1}\,|\sigma|}
\left[\int\limits_{c}^{b}\frac{dr}{\left(r-\lambda_{+}\right)R_1}+
\int\limits_{c}^{b}\frac{dr}{\left(r-\lambda_{-}\right)R_1}\right],
\end{equation}
where
\[\lambda_{\pm}=-\frac{\widetilde{E}+\widetilde{m}\mp
\sqrt{(\widetilde{E}+\widetilde{m})^2-4
\xi\sigma(1-2\lambda)}}{2\sigma\left(1-2\lambda\right)}.
\]
In the general case for the arbitrary values of $\xi$ and $\sigma$ the integrals $J_{1}$ and $J_{2}$ cannot be expressed in elementary functions. The carried out transformation of $J_{1}$ and $J_{2} $ is convenient because after replacement of the integration variable\cite{Beitmen}
\begin{equation}
r=\frac{c(b-d)-d(b-c)\sin^{2}\varphi}{b-d-(b-c)\sin^{2}\varphi}
\end{equation}
the integrals from right hand sides of the formulae (\ref{20}) and (\ref{21}) are transformed to the complete elliptic integrals of the first, second and third kind.\cite{part1,Prudnikov} As a result, we have obtained the transcendental equation
\begin{eqnarray}
\displaystyle&&\frac{2\sqrt{2\lambda-1}}{\sqrt{\left(a-c\right)\left(b-d\right)}}
\left\{\frac{|\sigma|\left(c-d\right)^2}{\bar{\Re}}\right.\left[\bar{N}_{1}F\left(\bar{\chi}\right)
+ \bar{N}_{2}E\left(\bar{\chi}\right)+
\bar{N}_{3}\Pi\left(\bar{\nu},\bar{\chi}\right)
\right.\nonumber \\
&&+ \left.\left.\bar{N}_{4}\Pi\left(\frac{d}{c}\,\bar{\nu},\bar{\chi}\right)\right]
+\displaystyle\frac{k}{2\left(2\lambda-1\right)|\sigma|}\left[(c-d)\left(\bar{N}_{5}
\Pi\left(\bar{\nu}_+,\bar{\chi}
\right)+\bar{N}_{6}\Pi\left(\bar{\nu}_-,\bar{\chi}\right)\right)\right.\right. \nonumber \\
&&\qquad\qquad\qquad\qquad\qquad\qquad\qquad\qquad\qquad\quad\left.\left.+
\bar{N}_{7}F\left(\bar{\chi}\right)\right]\right\}=\left(n_r+\frac{1}{2}\right)\pi,\label{eq22}
\end{eqnarray}
defining (in the quasi-classical approximation) the real part $E_r=E_{n_r k}$ of complex energy of quasistationary states at $U_{min}<\bar{E_r}<U_{max}$. Here the quantities $\bar{\nu}$,
$\bar{\nu}_{\pm}$, $\bar{\chi}$, $\bar{\Re}$, $\bar{\aleph}$,
$\bar{N}_j$ ($j=1,2,...,7$) are obtained from respective expressions (\ref{A.1})--(\ref{A.6}) for $\nu$, $\nu_{\pm}$, $\chi$, $\Re$,
$\aleph$, $N_j$ (recently found by us in Ref.~\refcite{part1})
by making the simultaneous replacements $a\rightarrow b$, $b\rightarrow c$,
$c\rightarrow d$ and $d\rightarrow a$.

For the determination of the Stark energy $E_r=E_{n_r k}$ the equation (\ref{eq22}) can be solved by numerical methods only. To construct the asymptotic behavior of resonances it is necessary to apply asymptotic methods of calculation of phase integrals just as in Sec.~4 of Ref.~\refcite{part1}. This imposes some restrictions in calculation of shifts of quasistationary levels and its widths for both small values of intensity $\sigma$ of radial-constant (scalar-vector) long-range field and not too large ones. Namely the quantity $\widetilde{m}$ divides the range of $\widetilde{E}_r=E_r-\lambda V_0$ into the two domains, in which the spectrum of quasistationary states has a various asymptotic behavior.
Consider some of the most typical situations connected with the relative values of the energy $\widetilde {E}_r$ and level
$\widetilde{m}$.

{\bf Case A}. If $\widetilde{E}_r<\widetilde{m}$, $U_{min}<\bar{E}_r<m$
(see Fig.~\ref{f1}, where $U_{min}=U(r_{min}, E_r)$,
$r_{min}\approx\gamma^2/\widetilde{E}_r\xi$) and condition
$\sigma/\xi \widetilde{m}^2\ll 1$ is satisfied, the pair of classical turning points $a$ and $d$ is rather distant from pair of points $c$ and $b$ (see. Ref.~\refcite{part1}):
\begin{equation}\label{root1}
a,b\approx\frac{\widetilde{E}_r\xi\pm \theta}
{\mu^2}\left[1-\frac{\widetilde{E}_r\xi\pm
\theta}{\mu^4}\left(\eta_1\pm\frac{\widetilde{m}\xi\eta_2}{\mu}\right)\sigma\right],
\end{equation}
\begin{equation}\label{root2}
c\approx-\displaystyle\frac{\widetilde{m}-\widetilde{E}_r}{\sigma}-\frac{\xi}{\widetilde{m}-\widetilde{E}_r},
\qquad
d\approx\displaystyle-\frac{\widetilde{m}+\widetilde{E}_r}{\sigma(1-2\lambda)}+\frac{\xi}
{\widetilde{m}+\widetilde{E}_r}.
\end{equation}
Hereafter, we use the notation
\[\theta=\sqrt{(\widetilde{E}_r k)^2-(\widetilde{m}\,\gamma)^2},\quad
\mu=\sqrt{\widetilde{m}^2-\widetilde{E}_r^2},\quad \eta_2=\lambda
\widetilde{m}+(1-\lambda)\widetilde{E}_r.
\]
In this case the derivation of asymptotic expansions of quantization integrals (\ref{20}), (\ref{21}) in a small parameter $\sigma/\xi \widetilde {m}^2$ is carried out just as in item A of Sec.~4 of Ref.~\refcite{part1} and gives the former expression for the level energy
\begin{equation}\label{eq15a}
E_r=\widetilde{E}_0+\lambda V_0+\frac{\sigma}{2\xi\widetilde{m}^2}
\left[\left(\frac{\xi^2\widetilde{m}^2}{\mu_0^2}-k^2\right)
\eta_{10}+\left(\frac{2\xi^2\widetilde{m}
\widetilde{E}_0}{\mu_0^2}-k\right)
\eta_{20}\right]+O\left(\left(\frac{\sigma}{\xi\widetilde{m}^2}\right)^2\right),
\end{equation}
\begin{equation}\label{eq15b}
\widetilde{E}_0=\widetilde{m}\left[1+\frac{\xi^2}{\left(n_r'+\gamma\right)^2}\right]^{-1/2},\quad
n_r'=n_r+(1+\mathrm{sgn}\,k)/2,
\end{equation}
and the quantities $\mu_{0}$, $\eta_{10}$ and $\eta_{20}$ are obtained from
$\mu$, $\eta_1$ and $\eta_2$ by replacing $\widetilde{E}_r$ by
$\widetilde{E}_0$.

The influence of weak radial-constant scalar and electric fields on the system of Coulomb levels has been analyzed before (see, for example Refs.~\refcite{part1,Lazur}). The analysis was carried out both on the basis of quasi-classical formulae (\ref{eq15a}) and by the numerical solution of the transcendental equation (\ref{eq22}). In particular, the calculations have shown that the sign change in $\sigma$ ($\sigma\rightarrow-\sigma$) leads to small changes of energy spectrum, if $|\sigma|\ll 0.2$ GeV$^2$.

{\bf Case B}. Let us now consider $\widetilde{m}<\widetilde{E}_r$ and
$m<\bar{E}_r<U_{max}$ (see Fig.~\ref{f1}, where $U_{max}=U(r_{max},
E_r)$ and $r_{max}\approx \eta_1 [(2\lambda-1)\sigma]^{-1}$). Here the ratio
$\sigma\gamma/\widetilde{E}^2_r$ plays a role of the small parameter. In this case the quasistationary states in the composite field (\ref{2}) exist only at the positive values of parameter $\sigma$.

Under the requirements formulated above, from (\ref{A2}) we have obtained
\begin{equation}\label{eq23}
 a\approx\displaystyle\frac{\widetilde{E}_r+\widetilde{m}}{\sigma(2\lambda-1)}
+\frac{\xi}{\widetilde{E}_r+\widetilde{m}},
\qquad
b\approx\displaystyle\frac{\widetilde{E}_r-\widetilde{m}}{\sigma}+\frac{\xi}
{\widetilde{E}_r-\widetilde{m}},\qquad
c,d\approx\frac{-\widetilde{E}_r\xi\pm \theta}
{\widetilde{E}_r^2-\widetilde{m}^2}.
\end{equation}
It is seen that turning points $c$ and $b$ are rather distant from one another ($a,b\gg c,|d|$). This allows to evaluate the quantization integral (\ref{3a}) analytically.

Further we shall give only the recipe of evaluation of the quantization integrals $J_{1,2}$. Just as in the item B of Sec.~4 of Ref.~\refcite{part1}, the integration range in (\ref{20}) and (\ref{21}) we divide into two domains $c\leqslant r\leqslant\widetilde{r}$ and $\widetilde{r}\leqslant r\leqslant b$ by introducing the dividing point $\widetilde{r}$ satisfying the requirement $c\ll\widetilde{r}\ll b$. In the first domain $c\leqslant r\leqslant\widetilde{r}$ we calculate the integrals (\ref{20}), (\ref{21}) by expanding the quasimomentum $p(r)$ in a power series in the parameters $r/a\ll 1$ and $r/b\ll 1$. In the second domain $\widetilde{r}\leqslant r \leqslant b$ the expansion of $p(r)$ we carry out in the small quantities $c/r\ll 1$ and $|d|/r\ll 1$.

When we add the asymptotic expansions of integrals over
$c\leqslant r \leqslant \widetilde{r}$ and
$\widetilde{r}\leqslant r \leqslant b$ the final result will not contain the quantity $\widetilde{r}$. So, we have obtained the equation
\begin{eqnarray}
\displaystyle\frac{\eta_1\sqrt{\widetilde{E}_r^2-\widetilde{m}^2}}
{2\sigma(2\lambda-1)}-\widetilde{\eta}\left(\frac{\eta_2^2}
{2\sigma(2\lambda-1)}+\lambda\,\xi\,\right)-\gamma
\arccos\left(\displaystyle\frac{-\widetilde{E}_r\xi}{\theta}\right)\qquad\qquad\qquad\quad
\nonumber\\
-\displaystyle\frac{\widetilde{E}_r\xi}{\sqrt{\widetilde{E}_r^2-\widetilde{m}^2}}
\log\left(\frac{\sigma\,\eta_2 \theta}
{4\,e\,(\widetilde{E}_r^2-\widetilde{m}^2)^2}\right)
-\displaystyle\frac{\mbox{sgn}\,k}{2}\,\arccos\left(\displaystyle\frac{-\widetilde{m}\xi}
{\theta}\right)=\left(n_r+\displaystyle\frac{1}{2}\right)\pi.\label{eq15}
\end{eqnarray}
If we expand the left hand side of (\ref{eq15}) in
$\widetilde{m}/\widetilde{E}_r\ll 1$ up to the terms of the third order, then we obtain the transcendental equation for the level energy $E_r$ which we solve by the method of consecutive iterations. Thus, we arrive at the expression for the energy (within $O(\sigma\gamma/\widetilde{E}_r^2)$)
\begin{eqnarray}
 E_r=\zeta^{-1}\Biggl\{B+
\Bigl(B^2+\zeta\Bigl[2\sigma(1-2\lambda)\Bigl(\xi
\log\frac{\sigma|k|(1-\lambda)}{
\left.4\widetilde{E}^{(0)}\right.^{2}}+3\xi+\lambda\xi \widetilde{A}+\pi N\Bigr)\Bigr.\Bigr.\Biggr.\nonumber\\
\Biggl.\Bigl.\Bigl.+\lambda\widetilde{m}^2(1-\lambda
\widetilde{A})\Bigr]\Bigr)^{1/2}\Biggr\}+\lambda V_0,\label{eq16}
\end{eqnarray}
where
\[
\widetilde{E}^{(0)}=E^{(0)}-\lambda
V_0,\quad\widetilde{A}=(2\lambda-1)^{-1/2}\log
\left[\left(1+\lambda+\sqrt{2\lambda-1}\right)(1-\lambda)^{-1}\right],
\]
\[\zeta=(1-\lambda)^2
A-\lambda-2\sigma\xi(1-2\lambda)/(\widetilde{E}^{(0)})^2,\,
B=(1-\lambda)(1-\lambda
\widetilde{A})\widetilde{m}-4\sigma\xi(1-2\lambda)/\widetilde{E}^{(0)},
\]
\[
N=\displaystyle
n_r+\frac{1}{2}+\frac{\mbox{sgn}\,k}{4}+\frac{1}{\pi}
\left(\gamma\arccos\left(-\frac{\xi}{|k|}\right)-\xi\right),
\]
\begin{equation}\label{eq24}
\widetilde{\eta}=(2\lambda-1)^{-1/2}\log \left[\left(\eta_1+
\sqrt{(2\lambda-1)(\widetilde{E}_r^2-\widetilde{m}^2})\right)\eta_2^{-1}\right],
\end{equation}
and $\widetilde{E}^{(0)}=E^{(0)}-\lambda V_0$, $E^{(0)}$ is the zeroth approximation of energy, by the choice of which the quantity $E_{n_r k}$ depends very weakly and in most cases one can take $E^{(0)}\approx E_r(\xi=0)$.

The expressions (\ref{eq15}), (\ref{eq16}) differ from the corresponding formulae in the case of purely discrete spectrum\cite{part1} only by replacements $\eta\rightarrow \widetilde{\eta}$ and $A\rightarrow
\widetilde{A}$. So, the substitution $\eta\rightarrow
\widetilde{\eta}$, $A\rightarrow \widetilde{A}$ transforms the equations $E^{WKB(as)}_{n_r k}$, obtained in the case of purely discrete spectrum, into the equations for a real part $E_r$ of energy (\ref{eq15}), (\ref{eq16}) of below-barrier resonances. Accuracy of the calculation of $E_r$ by the means of formulae (\ref{eq15}), (\ref{eq16}) is fully appropriate (see Sec.~4 of Ref.~\refcite{part1}), and usually there is no point to make the result more precise for practical purposes.

\section{Asymptotic coefficients of a wave function}

Asymptotic coefficients $C_{F,G}$ at the zero and $A_{F,G}$ at the infinity are the characteristic parameters of a wave function. We shall give simple analytical approximations for these coefficients which describe the results of numerical calculations quite well.

First of all we consider the construction rules of asymptotic expansions of solutions of the Dirac equation at zero ($r\rightarrow 0$) along with more standard expansions of solutions, when
$r\rightarrow\infty$. For the considered potentials (18) we have
\begin{equation}\label{9}
F,G=C_{F,G}\,r^{\gamma}+..., \quad r\rightarrow0,\qquad
C_{F}/C_{G}=(k-\gamma)/\xi,
\end{equation}
and for wave functions of the discrete spectrum (0$\leqslant\lambda <1/2$) the normalization condition
$\int\limits^{\infty}_{0}\left(F^2+G^2\right)dr=1$ is satisfied. Values of $F^2(0)$, $G^2(0)$ (or, more precisely, of $C^2_{F,G}$) define the probability to discover particles at small distances from one another and are of considerable physical interest especially in the case of systems, in which interactions of two various types (for example, the Coulomb interaction and long-range one $v(r)$) exist.

In the classically forbidden range $0\leqslant r<r_0$ the wave function of oscillating type (\ref{2a}) is changed by the solution decreasing exponentially with increasing $r$ (see Fig.~\ref{f1}).
Matching the WKB-solutions of the Dirac equation on both sides of the turning point $r_0$, for radial distribution functions $F(r)$ and $G(r)$ we obtain quasi-classical expressions in below-barrier range $r<r_0$:
\begin{eqnarray}\label{10}
 F(r)=(-1)^{n_r}\frac{C_{1}^{\pm}}{2}\left[\displaystyle\frac{E-V+m+S}{q(r)}\right]^{1/2}\exp
  \left[-\int\limits^{r_{0}}_{r}\left(q-\frac{k\,w}{q\,r}\right)dr\right],\\
 G(r)=\mathrm{sgn}\,k\,(-1)^{n_r}\frac{C_{1}^{\pm}}{2}\left[\displaystyle\frac{E-V-m-S}{q(r)}\right]^{1/2}
  \exp\left[-\int\limits^{r_{0}}_{r}\left(q-\frac{k\,\widetilde{w}}{q\,r}\right)dr\right]. \label{10a}
\end{eqnarray}
All integrals in (\ref{10}) and (\ref{10a}) are expressed through the quite complicated combination of the elliptic integrals. But in the cases $\widetilde {E}_r <\widetilde{m}$ and $\widetilde{E}_r>\widetilde{m}$ they can be calculated through elementary functions, using the relations $\sigma/\xi\widetilde{m}^2 \ll1$ and
$\sigma\gamma/\widetilde{E}_r^2 \ll1$ to expand the integrands into power series.

Let us first investigate the asymptotic behavior of the quasi-classical solutions (\ref{10}), (\ref{10a}) at $r\rightarrow 0$ for the lower levels ($\widetilde{E}<\widetilde{m}$) which are basically defined by the Coulomb potential ($\sigma/\xi\widetilde{m}^2 \ll1$).
Note that the larger the Coulomb parameter $\xi$, the smaller is the essential potential $v(r)$ at small distances. Before the evaluation of the asymptotic coefficients $C_{F,G}$ by means of formulae (\ref{10}), (\ref{10a}) it is necessary to expand the quasi-classical momentum $p(r)$ in potential $v(r)$. Then, using the technique of evaluation of the phase integrals from Sec.~2 and proceeding in (\ref{10}), (\ref{10a}) to the limit $r\rightarrow 0$, we obtain in zeroth approximation the expressions for the asymptotic coefficients at zero:
\begin{eqnarray}\label{12}
&\displaystyle\left|C_{F}\right|=\sqrt{\frac{\xi}{T\gamma}}\left(\frac{e
\theta_0}{2\gamma^2}\right)^{\gamma}\left[\frac{\theta_0\,(|k|-\gamma)}{\xi(\gamma\widetilde{m}+
|k|\widetilde{E}_0)}\right]^{\frac{\mathrm{sgn}\,k}{2}}\left(\frac{\xi\widetilde{E}_0+
\gamma\mu_0}{\theta_0}\right)^{\frac{\xi\widetilde{E}_0}{\mu_0}},\nonumber\\
&\displaystyle\frac{C_{F}}{C_{G}}=\frac{k-\gamma}{\xi}.
\end{eqnarray}
Here
$\theta_0=\sqrt{(\widetilde{E}_0\,k)^2-(\widetilde{m}\,\gamma)^2}$,
and the period of radial oscillations $T$ is given by the formula
\begin{eqnarray}\label{8b}
T\approx\frac{2\pi\xi\widetilde{m}^2}{\mu_0^3}.
\end{eqnarray}
When we derive the expression (\ref{12}) we use the quasi-classical requirement of the normalization (\ref{t}). Solving the Dirac equation (\ref{1}) at small distances (in range $0<r<b$) one can neglect the term with the linear potential ($\sigma=0$). Having used the asymptotic behavior of the normalized radial functions $F(r)$ and $G(r)$ of the relativistic Coulomb problem\cite{Ahiezer} at $r\rightarrow 0$ and the relations (\ref{9}), we find the more exact (than (\ref{12})) expression for $C_{F}$:
\begin{eqnarray}\label{15}
C_{F}^C=\frac{(2\mu_0)^{\gamma+1/2}}{\Gamma(2\gamma+1)}\left[\frac{(\widetilde{m}+\widetilde{E}_0)
\Gamma(2\gamma+n_r'+1)}{\frac{4\xi\widetilde{m}^2}{\mu_0}\left(\frac{\xi\widetilde{m}}{\mu_0}-k\right)n_r'!}\right]
^{1/2}\left(\frac{\xi\widetilde{m}}{\mu_0}-k-n_r'\right).
\end{eqnarray}
Here $n_r'=n_r+(1+\mathrm{sgn}\,k)/2$. The formulae (\ref{12}) and (\ref{15}) differ one from another within an error between the Stirling formula \[n!=\sqrt{2\pi}\exp\left\{(n+1/2)\log n-n\right\}(1+O(n^{-1}))\quad \mbox{at}\quad n\rightarrow\infty\] and the $\Gamma$-function.

For states with $\widetilde{E}_r>\widetilde{m}$, when the requirement $\sigma\gamma/\widetilde{E}_r^2\ll 1$ is satisfied, the Coulomb potential is essential only in the range of small distances and can be considered as a small perturbation in the basic range of particle localization (i.e. in classically allowed range $b<r<a$). This gives the possibility to exclude ($\sigma=0$) the linear part of the potential $v(r)$ from the quasi-classical momentum $p(r)$ when evaluating the integrals in exponents (\ref{10}), (\ref{10a}).
Then at $r\rightarrow 0$ the asymptotic behavior of radial wave functions $F(r)$ and $G(r)$ obtained in this way allows to determine the asymptotic coefficients:
\begin{eqnarray}\label{11}
\left|C_{F}\right|=\sqrt{\frac{\xi}{T\gamma}}\left(\frac{e
\theta}{2\gamma^2}\right)^{\gamma}\left[\frac{\theta\,(|k|-\gamma)}{\xi(\gamma\widetilde{m}
+|k|\widetilde{E}_r)}\right]^{\frac{\mathrm{sgn}\,k}{2}}
\exp\left[\frac{\xi\widetilde{E}_r}{\sqrt{\widetilde{E}_r^2-\widetilde{m}^2}}
\arccos\frac{\xi\widetilde{E}_r}{\theta}\right],
\end{eqnarray}
where quantity $\theta$ is defined in (\ref{root1}), and energy $\widetilde{E}_r$ is given by the formula (\ref{eq22}). Characteristic feature of the considered case is the fact that in the integral (\ref{t}), which defines a period of radial oscillations $T$, only the range of values of the integration variable $r$, where the Coulomb potential can be considered a perturbation, is essential. By neglecting the Coulomb interaction, we arrive at the expression
\begin{eqnarray}\label{8a}
T\approx\frac{2}{\sigma
(1-2\lambda)}\left[-\lambda\sqrt{\widetilde{E}_r^2-\widetilde{m}^2}
+(1-\lambda)\,\widetilde{\eta}\,\eta_2\right].
\end{eqnarray}

The asymptotic coefficients $A_{F}$, $A_{G}$ of radial wave functions at infinity are important physical parameters of bound states as well. Along with the coefficients $C_{F}$, $C_{G}$ at zero (\ref{9}), the asymptotic coefficients $A_{F,G}$ are continually encountered in quantum mechanics,\cite{Landau} atomic and nuclear physics,\cite{Smirnov,Yamabe} in the inverse problem of quantum scattering theory\cite{Newton,Vu} etc. For the potentials (\ref{2}) the quantities $A_{F,G}$ are related to asymptotic behaviors of the normalized radial wave functions by the relations
\begin{equation}\label{19}
F,G=A_{F,G}\,r^{\widetilde{\gamma}}
\exp\left(-\frac{\sqrt{1-2\lambda}\sigma}{2}\,r^2-
\frac{\eta_1}{\sqrt{1-2\lambda}}\,r\right),\quad \sigma
r\rightarrow\infty,
\end{equation}
where $\sigma>0$, $A_{F}=-\sqrt{1-2\lambda}\,A_{G}$,
$\widetilde{\gamma}=\displaystyle\frac{\eta_2^2}{2(1-2\lambda)^{3/2}\sigma}-
\frac{\lambda\xi}{\sqrt{1-2\lambda}}$, and the parameter $\lambda$
has values in the range $0\leqslant\lambda<1/2$.

In the below-barrier range $r>r_1=a$ far from the turning point $r_1=a$ under the requirements $\sigma/\xi\widetilde{m}^2\ll1$ and $\widetilde{E}<\widetilde{m}$, after the evaluation of integrals the quasi-classical solutions (\ref{2e}), (\ref{2f}) are of the form of decreasing exponents
\begin{eqnarray}\label{13}
&&\left(\begin{array}{ll}F\\G\end{array}\right)\approx\frac{1}{\sqrt{T
q_0}}\left(\begin{array}{ll} \sqrt{\widetilde{m}+\widetilde{E}_0+
(1-2\lambda)\sigma r}\\-\sqrt{\widetilde{m}-\widetilde{E}_0+\sigma
r}\end{array}\right)\left(\frac{4\mu_0^4\theta_0^{-1}
r}{\mu_0^2+\mu_0 q_0+\eta_{10}\sigma r}\right)^
{\frac{\xi\widetilde{E}_0}{\mu_0}}\nonumber\\
&&\times\left(\frac{\xi\widetilde{m}-k\mu_0}
{\xi\widetilde{m}+k\mu_0}\right)^{1/4}\left(\frac{\sqrt{1-2\lambda}q_0+
(1-2\lambda)\sigma
r+\eta_1+\xi\widetilde{E}(1-2\lambda)\sigma/\mu^2}{\sqrt{1-2\lambda}[\mu+
\xi(\lambda\mu^2+2\eta_1\widetilde{E})/\mu^3]
+\eta_1+\xi\widetilde{E}(1-2\lambda)\sigma/\mu^2}\right)^{\widetilde{\gamma}}\nonumber\\
&&\times\left(\frac{\xi\widetilde{E}_0-\gamma\mu_0}
{\xi\widetilde{E}_0+\gamma\mu_0}\right)^{\gamma/2}\exp\left[-\frac{q_0
r}{2}+\frac{\eta_1(\mu-q_0)}{2(1-2\lambda)\sigma}+
\frac{\xi\widetilde{E}(\mu+q_0)}{2\mu^2}\right],
\end{eqnarray}
where $q_0=\sigma\sqrt{(1-2\lambda)(r-c)(r-d)}$. The estimates show that by the satisfaction of the requirements $\sigma/\xi\widetilde{m}^2\ll 1$ and $\widetilde {E}<\widetilde{m}$ there is quite long range of distances $r$ which are much larger than size of the Coulomb hydrogen-like system ($r\gg\langle r\rangle$, see (37) or (38) in Ref.~\refcite{part1}) and much smaller than the distance $\widetilde{r}\approx(\widetilde{E}\xi/\eta_1\sigma)^{1/2}$ at which the Coulomb interaction becomes quantitatively comparable with the long-range interaction. In this range as the wave functions of zeroth approximation it is natural to take the unperturbed radial functions $F$ and $G$ of the relativistic Coulomb problem, and the potential $v(r)$ can be considered as a small perturbation.
Neglecting it, we arrive at the following quasi-classical expressions for $F$ and $G$
\begin{eqnarray}\label{13a}
&&\left(\begin{array}{ll}F\\G\end{array}\right)=\left(\begin{array}{ll}
\sqrt{\widetilde{m}+\widetilde{E}_0}\\-\sqrt{\widetilde{m}-\widetilde{E}_0}
\end{array}\right)A^{WKB}_Cr^
{\frac{\xi\widetilde{E}_0}{\mu_0}}e^{-\mu_0 r}
\nonumber\\
&&=\displaystyle\frac{1}{\sqrt{T \mu_0}}\left(\begin{array}{ll}
\sqrt{\widetilde{m}+\widetilde{E}_0}\\-\sqrt{\widetilde{m}-\widetilde{E}_0}
\end{array}\right)\left(\frac{\xi\widetilde{m}-k\mu_0}
{\xi\widetilde{m}+k\mu_0}\right)^{1/4}\left(\frac{\xi\widetilde{E}_0-\gamma\mu_0}
{\xi\widetilde{E}_0+\gamma\mu_0}\right)^{\gamma/2}\left(\frac{2\mu_0^2
r}{\theta_0}\right)^{\frac{\xi\widetilde{E}_0}{\mu_0}}e^{-\mu_0
r}.\nonumber\\
\,
\end{eqnarray}
Equating (\ref{13a}) to the asymptotic (at $r\rightarrow\infty $) representation of solutions of the Dirac equation in the Coulomb field\cite{Ahiezer}
\begin{eqnarray}\label{14}
\left(\begin{array}{ll}F\\G\end{array}\right)=
\left(\begin{array}{ll}
\sqrt{\widetilde{m}+\widetilde{E}_0}\\-\sqrt{\widetilde{m}-\widetilde{E}_0}
\end{array}\right)A_Cr^{\frac{\xi\widetilde{E}_0}{\mu_0}}e^{-\mu_0 r}
\end{eqnarray}
we obtain the explicit expression for the period of radial oscillations of the classical relativistic particle:
\begin{eqnarray}\label{17}
T=\frac{1}{2 \mu_0 |A_C|^2}\left(\frac{\xi\widetilde{m}-k \mu_0}{\xi\widetilde{m}+k
\mu_0}\right)^{1/2}\left(\frac{2e\mu_0^2}{\theta_0}\right)^{\frac{2\xi\widetilde{E}_0}{\mu_0}}
\left(\frac{\xi\widetilde{E}_0-\gamma\mu_0}{\xi\widetilde{E}_0+\gamma\mu_0}\right)^{\gamma}.
\end{eqnarray}
Here $A_C$ is the asymptotic coefficient of the Dirac radial wave functions in the Coulomb potential:
\begin{equation}\label{18} |A_C|=\left[\frac{(\xi\,\widetilde{m}-k\,\mu_0)\,\mu_0}{2\,\xi\,\widetilde{m}^2\,
\Gamma\left(2\gamma+n_r'+1\right)n_r'!}\right]^{1/2}
(2\mu_0)^{\frac{\xi\widetilde{E}_0}{\mu_0}}.
\end{equation}
Comparison of (\ref{13a}) and (\ref{14}) shows that their exponential and power factors are the same, however, the asymptotic coefficients $A^{WKB}_C$ and $A_C$ differ within an error between the Stirling formula and $\Gamma$-function.

The formula (\ref{13a}) is obtained by neglecting the linear part of the potential $v(r)$. This approximation was argued above by the means of the circumstance that under the quasi-classical requirements ($\sigma/\xi\widetilde{m}^2 \ll1$) there is a range of distances $\langle r\rangle\ll r\ll\tilde{r}$, in which the distortion of a wave function, caused by action of the linear part of the potential $v(r)$, can still be neglected and there is the law of decreasing radial wave functions (\ref{13a}) that is characteristic for the relativistic Coulomb problem. Change of the law of decreasement (\ref{13a}) of functions $F(r)$ and $G(r)$ to (\ref{19}) at $r\gg\tilde{r}$ appears because in EP $U(r,E)$ we have taken into account the quadratic (in $\sigma r$) terms which increase with increasing $r$ more rapidly than others and so play a role of the perturbation which destroys the asymptotic regime (\ref{13a}).  As a result of such an account, using the quasi-classical approximation (\ref{13}) for the normalized radial wave functions $F$ and $G$ at large $r$, we obtain the following expression for the asymptotic coefficient at infinity
\begin{eqnarray}\label{19b}
 A_{F}=2\mu_0 A_C(1-2\lambda)^{\widetilde{\gamma}+1/4}\sigma^{\widetilde{\gamma}}
\left(\frac{\mu_0^2}{\sigma}\right)^{\frac{\xi\widetilde{E}_0}{\mu_0}}
\left(\frac{\sqrt{1-2\lambda}\,\mu_0+\eta_{10}}{2}\right)^{-\frac{\xi\widetilde{E}_0}{\mu_0}-\widetilde{\gamma}}\nonumber\\
\times\exp\left[-\frac{(\sqrt{1-2\lambda}\,\mu-\eta_1)^2}{4(1-2\lambda)^{3/2}\sigma}\right].
\end{eqnarray}

We now proceed to the other limiting case $\sigma\gamma/\widetilde {E}^2\ll 1$ when the centrifugal potential $\gamma^2/2mr^2$ does not play an essential role at large distances and can be omitted in the quasi-classical momentum $p(r)=iq(r)$. Having expanded the quantity $q(r)=|p(r)|$ in powers of the Coulomb potential and calculated the integrals in exponents in (\ref{2e}), (\ref{2f}) under the requirement $\sigma\gamma/\widetilde{E}^2\ll1$, in the asymptotic domain $r\rightarrow\infty$ we arrive at formulae of type of (\ref{19}) for $F$ and $G$, in which
\begin{eqnarray}\label{19a}
 A_{F}&=&\frac{(1-2\lambda)^{1/4}}{\sqrt{T}}\left(\frac{2(1-2\lambda)\sigma}
{\eta_2}\right)^{\widetilde{\gamma}}
\exp\left[-\frac{2\eta_1^2-\eta_2^2}{4(1-2\lambda)^{3/2}\sigma}\right.\nonumber\\
&+&\left.\frac{\xi\widetilde{m}\eta_2}{2\sqrt{1-2\lambda}(\widetilde{E}^2-\widetilde{m}^2)}
+\frac{\xi\widetilde{E}}{\sqrt{\widetilde{E}^2-\widetilde{m}^2}}\arccos
\left(-\frac{\eta_1}{\eta_2}\right)\right],
\end{eqnarray}
and the period $T$ is determined by the previous expression (\ref{8a}).

\section{Width of quasistationary states}

The quasi-classical approximation (or WKB method) is the most often used method of the approximate solution of the relativistic quantum mechanical problems and leads to an obvious physical picture of percolation of a particle through the potential barrier in EP $U(r,E)$.

In the consideration above the quasistationary character of the Stark spectrum was ignored. Thereupon it is necessary to remind that for $1/2<\lambda\leqslant 1$ and any value of $\sigma\neq 0$, $U(r, E_r)$ has the shape of a potential with a barrier, owing to what there are quasistationary states with complex energy $E=E_r-i\Gamma/2$ instead of discrete levels. This is a result of the attraction of the term $-V^2/2$ which at large distances and $1/2 <\lambda\leqslant 1$ strongly suppresses the contribution of all other summands in (\ref{2}) and transforms the finite domain of the fermion motion into the infinite one. Thereby one can assert that the influence of the long-range vector field $V_{l.r.}(r)$, dominating at large distances, is manifested not only by the means of a modification of the energy spectrum of the system, but also leads to a non-zero probability of its decay due to a fermion passing through the potential barrier in EP $U(r,E_r)$.

The probability of tunnel transition of a particle from the bound state (with energy $E_r$) into the continuum state is defined by the imaginary part (i.e. by the width $\Gamma$) of complex energy of quasistationary states:\cite{Lazur}
\[
\Gamma=-2\,\mathrm{Im}[G^{*}(r)F(r)]_{r\rightarrow \infty}.
\]

Having calculated the flux of the particles outgoing to infinity by means of quasi-classical formulae (\ref{2k})-(\ref{2m}), we find the following expression for the level width $\Gamma$
\begin{equation}
{\rm\Gamma}=\frac{1}{T}\exp\left[-2\,\Omega\right],
\label{eq24a}
\end{equation}
\begin{eqnarray}
&T=\displaystyle 2\int\limits_c^b\frac{E_r-V}{p}\,dr,\quad
\Omega=\displaystyle \int\limits_b^a
\left(q-\frac{kw}{q\,r}\right)dr.& \label{eq25}
\end{eqnarray}

The obtained quasi-classical formula (\ref{eq24a}) is the relativistic generalization of the well-known Gamow formula for the width of a quasistationary level. The nontrivial moment of such a generalisation is the modification of expression for the period of oscillations $T$ and the occurrence of the additional factor in the preexponent of expression (\ref{eq24a}) that depends on a sign of the quantum number $k$ and is caused by the spin-orbit coupling in the mixture of the scalar $S(r)$ and vector $V(r)$ potentials.

Thus, in the quasi-classical approximation the problem is reduced to evaluation of two characteristic phase integrals $T$ and $\Omega$.

Having used again the notations of Sec.~4 of Ref.~\refcite{part1}, we write the quantity $q$ from (\ref{eq25}) in a form convenient for our purposes
\[q=|\sigma|\,\sqrt{2\lambda-1}\,\displaystyle
\frac{\sqrt{\left(a-r\right)\left(r-b\right)\left(r-c\right)\left(r-d\right)
}}{r},\]
where the parameter $\lambda$ ranges values from 1/2 up to 1. By an appropriate (similar to (\ref{20}), (\ref{21})) transformation of integrands the integrals in right hand sides of formulae (\ref{eq25}) are reduced to the complete elliptic integrals. Without detailing the aforementioned calculations of $T$ and $\Omega$, we give only the final result
\begin{eqnarray}\label{eq26a}
T&=&\frac{4}{|\sigma|\sqrt{\left(a-c\right)\left(b-d
\right)\left(2\lambda-1\right)}}\left\{d\widetilde{E}_r+\xi-\lambda
\sigma \left(d^2-\frac{\left(c-d\right)^2}
{2\left(1-\bar{\nu}\right)}\right)F\left(\bar{\chi}
\right)\right.  \nonumber \\
&+&\left.\frac{\lambda \sigma \bar{\nu} \left(c-d\right)^2}{2
\bar{\Re}}
E\left(\bar{\chi}\right)+\left(c-d\right)\left[\widetilde{E} -
\lambda \sigma
\left(2d+\frac{\left(c-d\right)\bar{\aleph}}{\bar{\Re}}\right)
\right]\Pi \left(\bar{\nu},\bar{\chi}\right)\right\},
\end{eqnarray}
\begin{eqnarray}\label{eq26b}
&&\Omega=\frac{2\sqrt{2\lambda-1}}{\sqrt{\left(a-c\right)\left(b-d\right)}}
\left\{-\frac{|\sigma|\left(b-c\right)^{2}}
{\Re}\Biggl[N_{1}F\left(\chi\right)+N_{2}E\left(\chi\right)+N_{3}
\Pi\left(\nu,\chi\right)+N_{4}\Biggr.\right.\nonumber\\
&&\left.\left.\times\Pi\left(\frac{c}{b}\nu,\chi\right)\right]+\frac{k}{2\left(2\lambda-1\right)|\sigma|}\left[(b-c)\left(N_{5}
\Pi\left(\nu_{+},\chi\right)+N_{6}
\Pi\left(\nu_{-},\chi\right)\right)+N_{7}F\left(\chi\right)\right]\right\}.\nonumber\\
\,
\end{eqnarray}
The quantities $\nu$,
$\nu_{\pm}$, $\chi$, $\Re$, $\aleph$, $N_j$ и $\bar{\nu}$,
$\bar{\nu}_{\pm}$, $\bar{\chi}$, $\bar{\Re}$, $\bar{\aleph}$,
$\bar{N}_j$ ($j=1,2,...,7$) belonging to (\ref{eq26a}), (\ref{eq26b}) are defined in (A.1)-(A.6) and
(\ref{eq22}).

The derived formulae (\ref{eq26a}), (\ref{eq26b}) are valid at large values of modules of phase integrals $T$, $\Omega$ and together with (\ref{eq24a}) solve the problem of calculation of width $\Gamma$ of Stark below-barrier resonances at
$U_{min}<\bar{E}_r<U_{max}$. However, these formulae are rather cumbersome and not too convenient for concrete calculations. With a purpose of deriving an analytical expression for the width of quasistationary level $\Gamma$ the calculations for the cases $\widetilde{E}_r <\widetilde{m}$ and $\widetilde{E}_r>\widetilde{m}$ should be carried out separately.

\begin{figure}[bp]
\centerline{\psfig{file=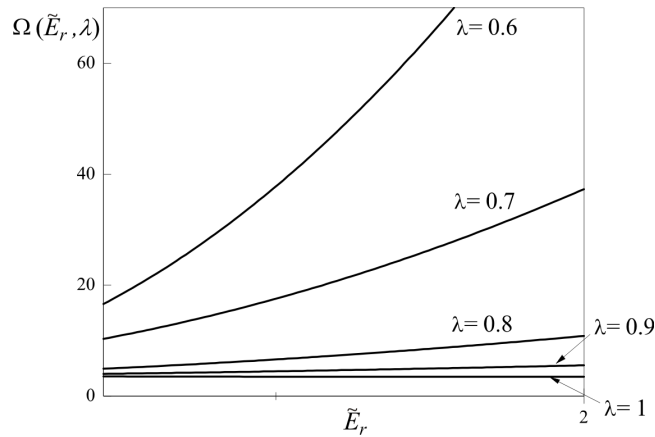,width=85mm}}
\vspace*{8pt}
\caption{The function $\Omega(E_r, \lambda)$ that defines the dependence of the exponential factor in the ionization probability (\ref{eq27}) on the level energy $E_r$, 0.44 GeV$<E_r<2$
GeV.}\label{Omega_Graph}
\end{figure}

{\bf Case A}. Let us begin with the simpler (in the sense of calculation) case of quasistationary levels with $\widetilde{E}_r>\widetilde{m}$ ($\bar{E}_r<U_{max}$) when the under requirements $\sigma\gamma\ll\widetilde{E}_r^2$, $\sigma>0$ the classical turning points $b$ and $a$ are rather distant from the pair of points $d$ and $c$. Asymptotic expansion of the barrier integral $\Omega$ can be constructed by means of the procedure which is very similar to the procedure applied to quantization integrals $J_1$ and $J_2$ in the case of purely discrete spectrum in the item A of Sec.~2. Omitting the details of the calculation, we give only the final formula for the width of the quasistationary level:
\begin{equation}
{\rm \Gamma }\approx\frac{1}{T}\exp\left[-2\Omega(\widetilde{E}_r,
\lambda)\right], \label{eq27}
\end{equation}
\begin{equation}\label{Omega}
\Omega(\widetilde{E}_r, \lambda)=\frac{\pi}{2\sqrt{2\lambda-1}}
\left(\frac{\eta_2^2}{\sigma(2\lambda-1)}+2\xi
\lambda+\frac{2\widetilde{E}_r\xi\sqrt{2\lambda-1}}
{\sqrt{\widetilde{E}_r^2-\widetilde{m}^2}}\right),
\end{equation}
where $\eta_2$ is defined above in (\ref{root1}).
Corresponding expansions for the energy $\widetilde{E}_r$ and period are given by previous formulae (\ref{eq15}), (\ref{eq16}) and (\ref{8a}). Analytical expressions (\ref{8a}), (\ref{eq27}) together with the known dependencies of $E_r(\sigma)$, obtained by means of the formulae (\ref{eq15}), (\ref{eq16}) or by numerical solution of the transcendental equation (\ref{eq22}), allow to calculate the widths of below-barrier resonances in ``transitional'' range of intensity  $\sigma$ of the radial-constant (scalar-vector) long-range field. As can be seen from the Fig.~\ref{Omega_Graph} the function $\Omega(E_r\lambda)$ decreases monotonically with increasing the parameter $\lambda$. Therefore, decreasing the relative weight $(1-\lambda)$ of the Lorentz-scalar $S_{l.r.}(r)$ in the long-range part $v(r)$ of the interaction (\ref{2}) rapidly increases a probability of ionization of quasistationary level.

When the level energy is close to the top of the barrier ($\bar{E}_r\rightarrow U_{max}$), the quasi-classical formula (\ref{eq27}) for the level width becomes invalid. From the point of view of the effective potential it corresponds to closing of the turning points $b$ and $a$ owing to what the exponential smallness of level width $\Gamma$ disappears. In this case in the below-barrier range $b<r<a $ EP $U(r,E_r)$ becomes parabolic, and the quantum mechanical problem, similar to calculation of penetrability of a parabolic barrier, arises. The formulae necessary for considering this case are given, for example, in papers.\cite{Mur3,Popov5}

{\bf Case B}. At $\widetilde{E}_r<\widetilde{m}$, $U_{min}<\bar{E_r}$
the asymptotic expansions of $T$ and $\Omega$ in positive powers of small dimensionless parameter $\sigma/\xi\widetilde{m}^2$ are constructed by the same technique, as in the item B of Sec.~2. In short, we shall derive only the asymptotic approximation of the integral $\Omega$ which defines the barrier factor. Assuming that the potential barrier in EP $U(r, E_r) $ is rather wide, we divide the integration domain $b\leqslant r\leqslant a $ into two segments by the point $r^*$ which satisfies the condition $b\ll r^*\ll a$ (it is indeed possible because $a\rightarrow\infty$ as $\sigma\rightarrow 0$). In the first domain $b\leqslant r\leqslant r^*$, the long-range potential $v(r)$ can be considered as a small perturbation, and in the second one $r^*\leqslant r\leqslant a$, on the contrary, the Coulomb field is much weaker than the long-range field and can be considered as a small perturbation. Expanding the quantity $q(r)$ in small perturbation in each domain, we obtain some tabular integrals in (\ref{eq25}), the sum of which gives the value of the barrier integral $\Omega$ to within terms $O(\sigma/\xi\widetilde{m}^2)$. By omitting the details, sense of which is clear, we give the asymptotic behavior of the imaginary part of energy of quasistationary state in a weak-coupling regime:
\begin{eqnarray}
\Gamma\approx\frac{1}{T}\left(\frac{4\mu_0^4 e}{|\sigma|\eta_{20}
\theta_0}\right)
^{\frac{2\widetilde{E}_0\xi}{\mu_0}}\left(\frac{\widetilde{m}\,\xi-\mu_0\,
k}{\widetilde{m}\,\xi+\mu_0\,k}\right)^{1/2}\left(\frac{\widetilde{E}_0\,\xi-\mu_0\,\gamma}
{\widetilde{E}_0\,\xi+\mu_0\,\gamma}\right)^{\gamma}\qquad\qquad\qquad\qquad\nonumber\\
\times\exp\left\{-\frac{\mu\eta_1}{\sigma\,(2\lambda-1)}
-\frac{1}{|\sigma|\sqrt{2\lambda-1}}\left(\frac{\eta_2^2}{2\lambda-1}+2\lambda\,\xi\,\sigma\right)
\arccos\left(-\frac{\eta_1}
{\eta_2}\,\mbox{sgn}\,\sigma\right)\right\}. \label{eq29}
\end{eqnarray}
This result is valid for both positive and negative values of parameter $\sigma$, and in the potential well $c<r<b$ one can again use the formula (\ref{8b}) or (\ref{17}) for the period of oscillations of a classical relativistic particle with energy $E_r$. If one uses more exact expression for the period (\ref{17}), derived by matching WKB solution (\ref{13a}) with the asymptotic behavior of the relativistic Coulomb wave function at $r\rightarrow\infty $, then one can represent the width of quasistationary levels (\ref {eq29}) in the form
\begin{eqnarray}
\Gamma&=&2\mu_0\left|A_C\right|^2\left(\frac{2\mu_0^2}{|\sigma|\eta_{20}}\right)
^{\frac{2\xi\widetilde{E}_0}{\mu_0}}
\exp\left\{-\frac{\Phi(\widetilde{E}_0,
\lambda)}{|\sigma|}-\frac{2\lambda\mu_0\rho}{2\lambda-1}\right.\nonumber\\
&-&\left.\frac{2\,\mbox{sgn}\,\sigma}{\sqrt{2\lambda-1}}\left[\frac{(1-\lambda)\eta_{20}\rho}{2\lambda-1}+\lambda\xi\right]
\arccos\left(-\,\frac{\eta_{10}}
{\eta_{20}}\mbox{sgn}\,\sigma\right)\right\}, \label{eq30}
\end{eqnarray}
where $A_C$ is the asymptotic coefficient (\ref{18}) of the normalized wave function in the Coulomb potential, and notations
$\widetilde{E}_0$, $\mu_0$, $\eta_{10}$ and $\eta_{20}$ were introduced in (\ref{eq15a}), (\ref{eq15b}). The functions $\Phi(\widetilde{E}_0, \lambda)$ and $\rho(\widetilde{E}_0, \lambda)$ from the exponent (\ref{eq30}) are given by the formulae
\begin{equation}\label{Phi}
\Phi(\widetilde{E}_0,
\lambda)=(2\lambda-1)^{-1}\left\{\frac{\eta_{20}^2}{\sqrt{2\lambda-1}}\,
\arccos\left(-\frac{\eta_{10}}
{\eta_{20}}\mbox{sgn}\,\sigma\right)+\eta_{10}\mu_0\,
\mbox{sgn}\,\sigma\right\},
\end{equation}
\[
\rho(\widetilde{E}_0, \lambda)=\frac{1}{2\xi\widetilde{m}^2}
\left[\left(\frac{\xi^2\widetilde{m}^2}{\mu_0^2}-k^2\right)
\eta_{10}+\left(\frac{2\xi^2\widetilde{m}
\widetilde{E}_0}{\mu_0^2}-k\right) \eta_{20}\right].
\]
\begin{figure}[bp]
\centerline{\psfig{file=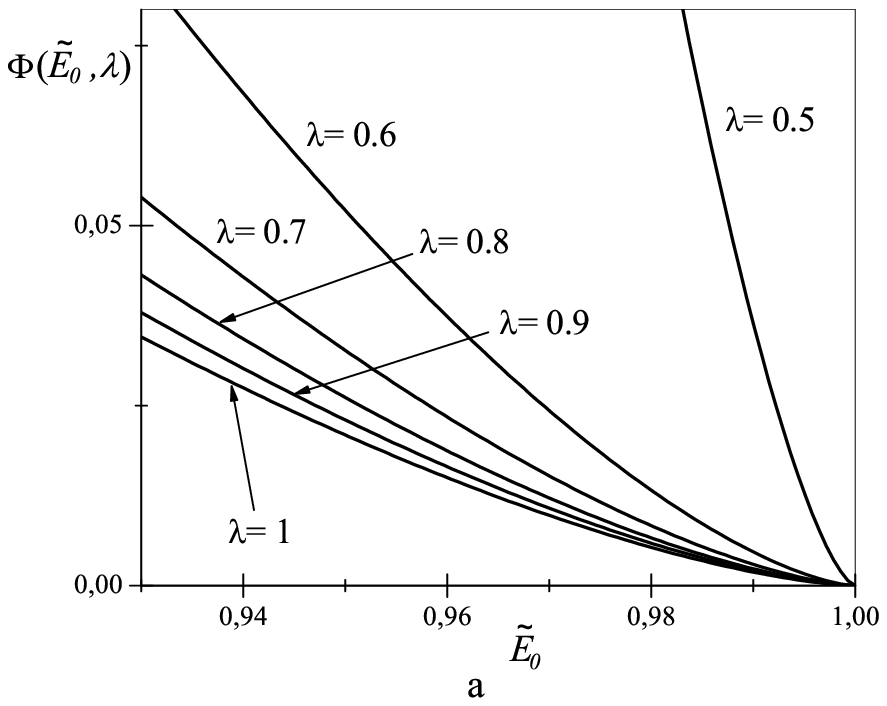,width=85mm}}
\centerline{\psfig{file=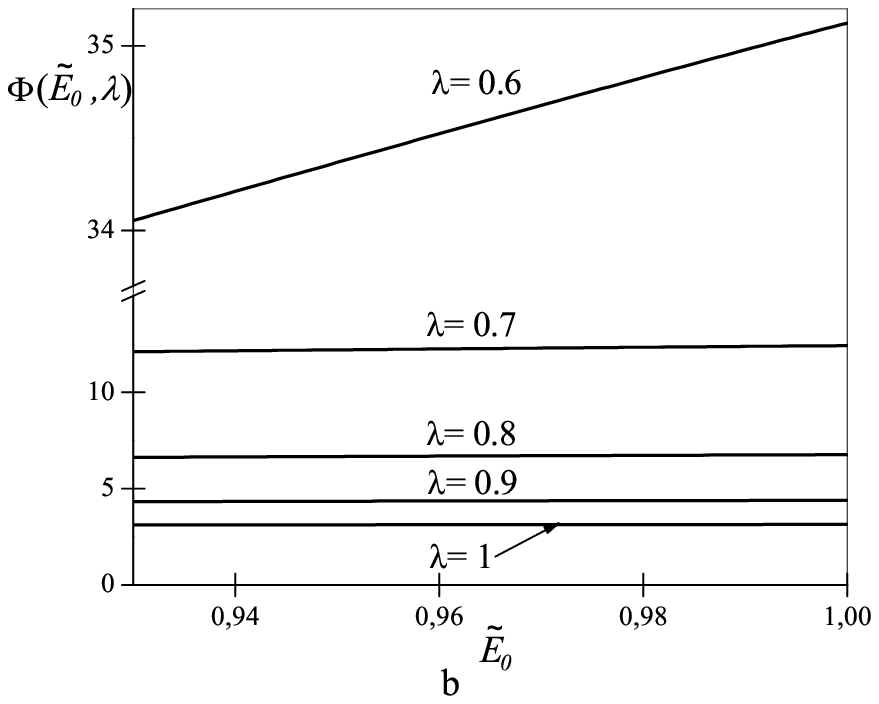,width=85mm}}
\vspace*{8pt}
\caption{The function $\Phi(\widetilde{E}_0, \lambda)$ that defines the dependence of the exponential factor in the ionization probability (\ref{eq30}) on the level energy $E_0$ (in GeV): a) for $\sigma<0$, b) for $\sigma>0$.}\label{fig_Phy}
\end{figure}

Comparison of the results of calculations of level energies based on the ``exact'' quasi-classical formulae (\ref{eq24a}), (\ref{eq26a}), and (\ref{eq26b}) with results of numerical calculations shows that
relative error of (\ref{eq30}) and (\ref{Phi}) does not exceed 2$\%$ at $|\sigma|\sim 10^{-6}-10^{-4}$ GeV$^2$.
So the formula (\ref{eq30}) is convenient when used for quick estimates of $\Gamma$. In Fig.~\ref{fig_Phy} the dependence of the function $\Phi(\widetilde{E}_0,\lambda)$ on level energy $\widetilde{E}_0$ are represented for several values of mixing parameter $\lambda$. As is shown in Fig. (\ref{fig_Phy}a, $\sigma <0$), $\Phi(\widetilde {E}_0,\lambda)$ increases with sinking of the level $\widetilde{E}_0$ and decreases when the mixing coefficient $\lambda$ ($1/2<\lambda\leqslant 1$) increases.
The last fact can be easily explained: by increasing the relative weight $\lambda$ (from 1/2 up to 1) of the Lorentz-vector $V_{l.r.}(r)$ in the long-range part of $U(r, E_0)$ the attraction increases at large distances and the effective width of a barrier decreases. It results in the fact that at $|\sigma |\sim 10^{-6}-10^{-4}$ GeV$^2$ the increasing of the parameter $\lambda$ ($1/2 <\lambda\leqslant 1$) leads to the increasing of the ionization probability $\Gamma$ and, on the contrary, decreases $\widetilde{E}_0$ or, in other words, the sinking of the bound level decreases $\Gamma$ (at the fixed values $\lambda$ and $\sigma$). In the case of the positive values of $\sigma$ the function $\Phi(\widetilde{E}_0,\lambda)$ decreases monotonically with increasing the parameter $\lambda$ (Fig.~\ref{fig_Phy}), therefore by decreasing the relative weight $(1-\lambda)$ of the Lorentz-scalar $S_{l.r.}(r)$ in the long-range part $v(r)$ of the interaction (\ref{2}) one rapidly increases the probability of ionization of quasistationary level (at the same value of $\sigma$).

The formula (\ref{eq30}) is asymptotically exact at the limit
$\sigma\rightarrow 0$. At small $\sigma\neq 0$ only the exponential factor, depending extremely rapidly on intensity of a long-range field and varying within many orders of value, plays an essential role in (\ref{eq30}). However, the applicability field of the formula (\ref{eq30}) is rather narrow: $|\sigma|\lesssim 10^{-4}$
GeV$^2$.

As can be seen from (\ref{eq30}), the width of quasistationary level is proportional to the squared asymptotic coefficient
$A_C$. It is no wonder: at $\sigma/\xi\widetilde{m}^2$ $\ll 1$ and $\widetilde{E}_r<\widetilde{m}$ the ionization goes from the ``tail'' of the Coulomb wave function and the barrier is wide.

In the case when in addition to the Coulomb field there is only the radial-constant electric field ($\lambda=1$), from the formula (\ref{eq30}) at $V_0=0$, $\sigma<0$ and $\xi=Z\alpha$ ($Z$ is the nuclear charge, $\alpha\approx1/137$ is the fine structure constant) we have obtained ($\hbar=m=c=1$):
\begin{eqnarray}
\Gamma=2\mu_0\left|A_C\right|^2
\left(\frac{2\mu_0^2}{|\sigma|}\right)
^{\frac{2E_0Z\alpha}{\mu_0}}
\exp\left[-\frac{\Phi(E_0)}{|\sigma|}+2Z\alpha\arccos
E_0-2\rho\sqrt{1-E_0^2}\right], \label{eq31}
\end{eqnarray}
where $E_0$ is the energy of a bound state in the absence ($\sigma=0$) of an external long-range field, and
\[
\rho=\frac{1}{2Z\alpha}\left[E_0\left(\frac{3Z^2\alpha^2}{1-E_0^2}-k^2\right)-k\right].
\]
The function $\Phi(E_0)$ in the exponent is given by the expression
\begin{eqnarray}
\Phi(E_0)=\arccos E_0-E_0\sqrt{1-E_0^2}, \label{eq32}
\end{eqnarray}
and possesses the obvious property $\Phi(-E_0)=\pi-\Phi(E_0)$.

Consider some of the limiting cases of the derived expression
(\ref{eq31}):
\begin{enumerate}
\item Let us begin with ionization of $s$-level bound by short-range ($Z=0$) forces under the influence of the radial-constant electric field $\sigma<0$. In this case from (\ref{eq31}) we obtain the expression
\begin{eqnarray}
\Gamma \propto \exp\left[-\frac{\Phi(E_0)}{|\sigma|}\right],
\label{eq33}
\end{eqnarray}
which coincides with the result of papers\cite{Mur_MMB} for the Stark ionization of $s$-level bound by short-range potential or $\delta$-potential (which is a good approximation in the case of ionization of the single charged negative ions, such as $H^{-}$, $Na^{-}$ etc.) within exponential accuracy.

\item In the presence of the Coulomb field it is worthwhile to consider the various limiting cases for the quantities appearing in the formula (\ref{eq31}):
\begin{equation}\label{eq34}
\arccos
E_0=\left\{\begin{array}{cc}\displaystyle\left(1-E_0^2\right)^{1/2}+\frac{1}{6}\left(1-E_0^2\right)^{3/2}+...,&
E_0\rightarrow1,\\\displaystyle\frac{\pi}{2}-E_0-\frac{1}{6}E_0^3+...,&E_0\rightarrow0,
\\ \displaystyle\pi-\left(1-E_0^2\right)^{1/2}-\frac{1}{6}\left(1-E_0^2\right)^{3/2}+...,&
E_0\rightarrow-1,
\end{array}\right.
\end{equation}
\begin{equation}\label{eq35}
\Phi(E_0)=\left\{\begin{array}{cc}\displaystyle\frac{2^{5/2}}{3}(1-E_0)^{3/2}\left[1-\frac{3}{20}(1-E_0)+...\right],&
E_0\rightarrow1,\\\displaystyle\frac{\pi}{2}-2E_0+\frac{1}{3}E_0^3+...,&E_0\rightarrow0,
\\ \displaystyle\pi-\frac{2^{5/2}}{3}(1+E_0)^{3/2}+...,&
E_0\rightarrow-1.
\end{array}\right.
\end{equation}
\end{enumerate}

In the nonrelativistic limit ($E_0\rightarrow1$, $ \alpha\rightarrow0$) the formula (\ref{eq31}) transforms into the known expression\cite{Mur_Popov} for width of below-barrier resonances in spherical model of the Stark effect for hydrogen atom:
\begin{eqnarray}
\Gamma=2\mu_{n.r.}\left|A_{n.r.}\right|^2
\left(\frac{2\mu_{n.r.}^2}{|\sigma|}\right)^{2n}
\exp\left(-\frac{2\mu_{n.r.}^3}{3|\sigma|}\right),
\label{eq36}
\end{eqnarray}
where $\mu_{n.r.}=Z/n$, $n$ is the principle quantum number,
$A_{n.r.}$ is the asymptotic coefficient (at infinity) of the Coulomb wave function of the free ($\sigma=0$)
nonrelativistic hydrogen-like atom.

The correction of the order of $\alpha$ in the exponent of (\ref{eq31}) slightly increases the probability of ionization when compared with the corresponding nonrelativistic formula (\ref{eq36}). The factor $\Phi (E_0)$ in the exponent of (\ref{eq31}) increases monotonically with sinking level (it is equal to $\pi/2$ and $\pi $ at $E_0=0$ and $E_0 =-1$, respectively) which leads to rapid decreasing the probability of ionization.

At $E_0\rightarrow-1$, that is for the level which has sunk to the edge of the negative energy continuum, the leading (exponential) factor in (\ref{eq31}) becomes equal to $\exp(-\pi / |\sigma |)$ and coincides with the corresponding factor in the Schwinger formula\cite{Schwinger} (obtained within the quantum field theory) for the probability of creation of the electron-positron pairs from vacuum in the constant electric field.

By means of the formulae obtained above the spectrum of quasistationary levels is described for the accepted hybrid version of SMSE. Such model qualitatively reproduces the following characteristic features of quasistationary states in an mixture of scalar and vector potentials of barrier type (\ref{2}): 1) very strong (at small $\sigma$) dependence of $\Gamma$ on the binding energy of tunneling fermion and on the mixing coefficient $\lambda$; 2) nonanalytic dependence of shift and width of level on the ``force'' $\sigma$ of scalar and vector long-range interactions.

In conclusion, note that the version of SMSE considered above could seem to be rather artificial and having no relation to real problems. Let us emphasize thereupon that the Dirac equation with the potentials (\ref{2}) at $1/2<\lambda\leqslant1$ can serve as the etalon equation for the relativistic theory of quasistationary states with the scalar-vector variant of interactions. As is known, now there are all grounds to consider that such interactions exist between composite objects (quarks and gluons) of QCD. Also, in the nuclear reactions of the tunnel type the peculiar features of a scalar-vector variant of interactions, that should be taken into account when calculating the penetration probability of tunneling fragments through potential barriers, can be fully manifested.

Examples of strong influence of a scalar field on spectra of resonance states of strongly interacting particles are given in the recent analysis (see, for example review\cite{Anisovich}) of experimental data from the Crystal Barrel Collaboration on the in-flight proton-antiproton annihilation into mesons of the final state below 2400 MeV. In particular, in the paper\cite{Anisovich} the following assumption has been made. The existence of light $\sigma$-meson can be caused by singular behavior of quark-antiquark interaction $\sim 1/q^{4}$ at the small transmitted momentums. In the co-ordinate space it corresponds to linear increasing of the potential $v(r)$ at large distances.

\appendix
\section{Some notations}

The quantities introduced in Sec.~2 have the form
\begin{equation}\label{A.1}
\nu=\displaystyle\frac{a-b}{a-c},\quad \nu_{\pm}=
\frac{\lambda_{\pm}-c}{\lambda_{\pm}-b}\,\nu, \quad \chi=\sqrt{\nu
\frac{(c-d)}{(b-d)}},\quad
\Re=\left(1-\nu\right)\left(\chi^{2}-\nu\right),
\end{equation}
\begin{equation}\label{A.2}
N_{1}=\frac{\chi^{2}\left(b-c\right)}{4}- \frac{3\aleph
\left(b-c\right)}{8\left(1-\nu\right)}-\frac{(\chi^{2}-\nu)(f+3c)}{2}
+\frac{\Re(c^{3}+c^{2}f+cg+h+l/c)}{\left(b-c\right)^{2}},
\end{equation}
\begin{equation}\label{A.3}
N_{2}=-\frac{\nu}{2}\left[f+3c+\frac{3}{4}\frac{\left(b-c\right)\aleph}
{\Re}\right],
\end{equation}
\begin{eqnarray}\label{A.4}
N_{3}=\frac{1}{2}\left\{\frac{3\left(b-c\right)\aleph^{2}}{4\Re}+\frac{2\Re(3c^{2}+2cf+g)}{\left(b-c\right)}
+\left(b-c\right)\left[\left(1+\chi^{2}\right)\nu-3\chi^{2}\right]\right.\nonumber\\
+\left.\aleph\left(f+3c\right)\right\},
\end{eqnarray}
\begin{equation}\label{A.5}
N_{4}=-\frac{\Re}{\left(b-c\right)}\frac{l}{b c},\quad
N_{5}=[(b-\lambda_{+})(\lambda_{+}-c)]^{-1},\quad
N_{6}=[(b-\lambda_{-})(\lambda_{-}-c)]^{-1},
\end{equation}
\begin{equation}\label{A.6}
N_{7}=\frac{2}{(\lambda_{+}-c)
(\lambda_{-}-c)}\left(c+\frac{\widetilde{E}+\widetilde{m}}{2(1-2\lambda)\sigma}\right),
\quad \aleph=\chi^{2}\left(3-2\nu\right)+\nu\left(\nu-2\right).
\end{equation}

\end{document}